\documentclass[conference]{IEEEtran}

\usepackage[normalem]{ulem}
\usepackage{cite}
\usepackage{algorithm, algorithmic, lipsum}
\usepackage{amsmath,amssymb, amsfonts, dsfont, braket}
\usepackage{mathtools, braket, lipsum, bibentry}
\usepackage{graphicx, multirow, tabularx, hyperref}
\usepackage{textcomp}
\usepackage[usenames,dvipsnames]{xcolor}
\usepackage[utf8]{inputenc} 
\usepackage[T1]{fontenc}    
\usepackage{booktabs}       
\usepackage{nicefrac}       
\usepackage{microtype}      
\usepackage{float}
\usepackage{enumitem, lipsum}
\usepackage[caption=false,font=footnotesize]{subfig}

\def\BibTeX{{\rm B\kern-.05em{\sc i\kern-.025em b}\kern-.08em
    T\kern-.1667em\lower.7ex\hbox{E}\kern-.125emX}}
\graphicspath{{../figures/}{figures/}}

\newcommand{\xrm}{\textrm{x}}

\hypersetup{colorlinks=true, linkcolor=black, urlcolor=blue}

\graphicspath{{../figures/}{figures/}}

\begin{document}
\title{Adaptive mitigation of time-varying quantum noise}

\author{
\IEEEauthorblockN{Samudra~Dasgupta$^{1, 2^*}$, Arshag~Danageozian$^{3^{\#}}$ and Travis S.~Humble$^{1,2^\dagger}$\\\\}
\IEEEauthorblockA{
$^1$Quantum Science Center, Oak Ridge National Laboratory, Oak Ridge, Tennessee, USA\\
$^2$Bredesen Center, University of Tennessee, Knoxville, Tennessee, USA\\
$^3$ Hearne  Institute  for  Theoretical  Physics,  Department  of  Physics  and  Astronomy,\\
Louisiana  State  University,  Baton  Rouge,  Louisiana  70803,  USA\\\\
$^*$sdasgup3@tennessee.edu, ORCID: 0000-0002-7831-745X\\
$^\#$arshag.danageozian@gmail.com, ORCID: 0000-0003-0044-9951\\
$^\dagger$humblets@ornl.gov, ORCID: 0000-0002-9449-0498\\
}}

\maketitle


\begin{abstract}
Current quantum computers suffer from non-stationary noise channels with high error rates, which undermines their reliability and reproducibility. We propose a Bayesian inference-based adaptive algorithm that can learn and mitigate quantum noise in response to changing channel conditions. Our study emphasizes the need for dynamic inference of critical channel parameters to improve program accuracy. We use the Dirichlet distribution to model the stochasticity of the Pauli channel. This allows us to perform Bayesian inference, which can improve the performance of probabilistic error cancellation (PEC) under time-varying noise. Our work demonstrates the importance of characterizing and mitigating temporal variations in quantum noise, which is crucial for developing more accurate and reliable quantum technologies. Our results show that Bayesian PEC can outperform non-adaptive approaches by a factor of 4.5x when measured using Hellinger distance from the ideal distribution. 
\end{abstract}

\begin{IEEEkeywords}
Device reliability, 
Computational accuracy, 
Result reproducibility, 
Probabilistic error cancellation,
Adaptive mitigation,
Spatio-temporal non-stationarity,
Time-varying quantum noise,
NISQ hardware-software co-design
\end{IEEEkeywords}

\section{Introduction}
Characterizing noise in multi-qubit quantum devices is a fundamental step towards achieving fault-tolerant quantum computation. However, there is an increasing body of evidence showcasing that physical platforms for quantum computation are subject to both spatial and temporal variations \cite{muller2015interacting, klimov2018fluctuations}. Namely, different qubit locations in a given quantum device experience different noise profiles, and furthermore, the statistical characteristics of the noise vary over time. These have important implications for current and future quantum technologies, especially when quantum tasks are performed on time/length scales larger than the characteristic time/length of the noise variations. 
An adaptive treatment of these complications is of direct relevance to quantum computing.
\par 
Modern (classical) computers have device components with failure rates of $10^{-17}$ or less. Current state-of-the-art quantum computers however exhibit gate-level (e.g. CNOT) failure rates of around $10^{-2}$ \cite{bravyi2021mitigating}. Consequently, reliability and reproducibility of results obtained from quantum computers is a problem in modern quantum information science \cite{kliesch2021theory, coveney2021reliability, baker20161, nichols2021opinion, blume2010optimal, ferracin2021experimental}. The problem is made worse by our experimental observation that today's NISQ (Noisy Intermediate-Scale Quantum) \cite{preskill2019quantum} devices are unstable. 
Also, the calibration may not be up-to-date due to device drift \cite{proctor2020detecting, znidaric2004stability, zhang2021predicting}. Also, correlations within a circuit can increase errors in one part while reducing errors in another. Thus, it is often not clear what is the right optimal operating point for the compensated circuit. 
\par
Key device characterization metrics often fluctuate with time in-between calibrations (temporal instability) and also across the chip (spatial stability). These types of time-varying noise require mitigation methods that operate between calibration periods and adapt to the evolving noise sources. The quantum noise channel in fact is a random variable that needs adaptive treatment. Using inaccurate characterization data for mitigation can impact circuit reproducibility \cite{proctor2020detecting}. 
\par 
Despite rapid developments, physical platforms for quantum computation exhibit temporal variations in noise parameters. For example, temporal noise variations in superconducting qubits are among the most well-studied in literature. The current consensus attributes the variations in the latter to the presence of certain oxides on the superconductors' surface, modeled as fluctuating two-level systems (TLS) \cite{muller2015interacting, klimov2018fluctuations}. Unlike temporal variations in noise, the effects of which are magnified by the complexity of a quantum circuit, the spatial variations are dependent on the geometry and scale of the circuit implementation. 
\par 
There exists an active research area that aims at addressing the time-varying nature of quantum noise within our current technology. This includes modeling \cite{etxezarreta2021time} and characterizing such noise sources, tracking their temporal profile \cite{bonato2017adaptive, cortez2017rapid, proctor2019detecting, wu2021continuous, danageozian2022noisy, turner2022spin}, as well as incorporating this statistical knowledge into existing quantum architectures via novel techniques, such as ``spectator systems'' and feedback control \cite{majumder2020real, danageozian2022recovery, song2023optimized}, continuous control \cite{kwon2022reversing}, and autonomous control \cite{lebreuilly2021autonomous}, among many others. Similarly, many advances have also been made for spatially varying noise in quantum devices \cite{klesse2005quantum}, its effect on the choice of circuit geometry \cite{gupta2020integration}, as well as the interplay with cross-talk between qubits for single and two-qubit gates \cite{parrado2021crosstalk, fang2022crosstalk}. 
\par 
In this article, we provide a spatio-temporal characterization of the noise variations of a transmon based IBM quantum computer. We model the multi-qubit noise by the twirled amplitude and phase damping channels. Then, we measure and characterize the variations in single qubit $T_{1}$ and $T_{2}$ times on various relevant time and length scales. Furthermore, we show how to infer the twirled channel parameters dynamically from the noisy binary output of the quantum circuit execution. Finally, we propose a Bayesian technique \cite{lukens2020practical, zheng2020bayesian, gordon1993novel, kotecha2003gaussian} to implement noisy gates using probabilistic error cancellation \cite{temme2017error, endo2018practical} in the presence of spatio-temporal variations of the twirled channels, and show how it outperforms current noisy implementations, without taking into account the spatio-temporal dependence. Our framework could be generalized and applied to various other techniques of adaptive error correction and mitigation, thereby minimizing the effects of spatio-temporally varying noise during quantum computation.
\par
The manuscript is organized as follows. In Section~\ref{sec:decoherence}, we review evidence for spatio-temporally varying noise in superconducting qubits. Section~\ref{sec:model} describes our theoretical model, along with the quantification of the degree of non-stationarity of the noise. Section~\ref{sec:bayesian} sets the adaptive Bayesian framework for estimating the non-stationary noise parameters, which we then use to perform probabilistic error cancellation to achieve high-fidelity qubit gates. Finally, we provide the conclusion in Section~\ref{sec:conc}, along with open questions and future directions.
\section{Stochastic fluctuation in register decoherence}\label{sec:decoherence}
\begin{figure}
\centering
\includegraphics[width=.84\columnwidth]{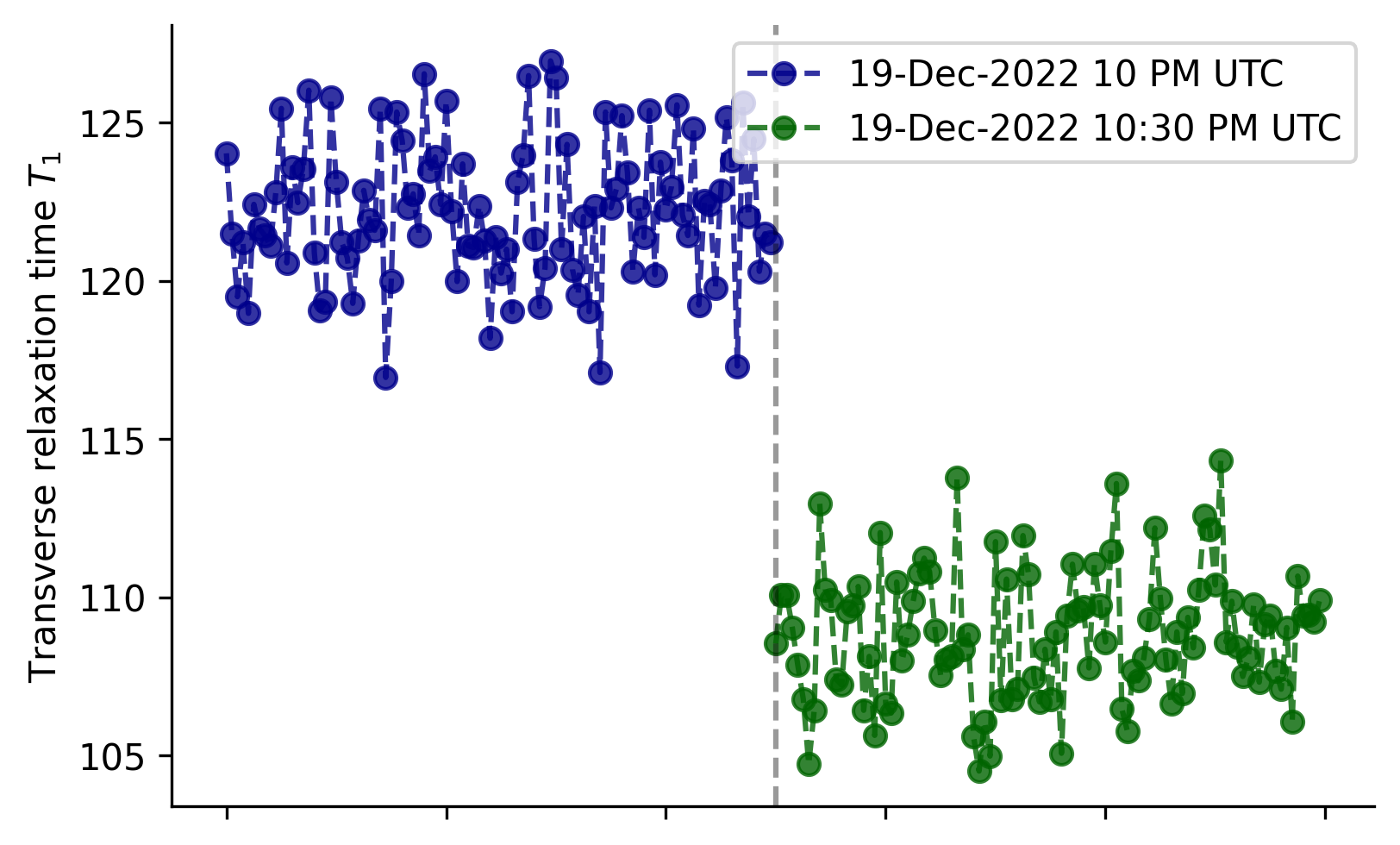}\\
\includegraphics[width=.84\columnwidth]{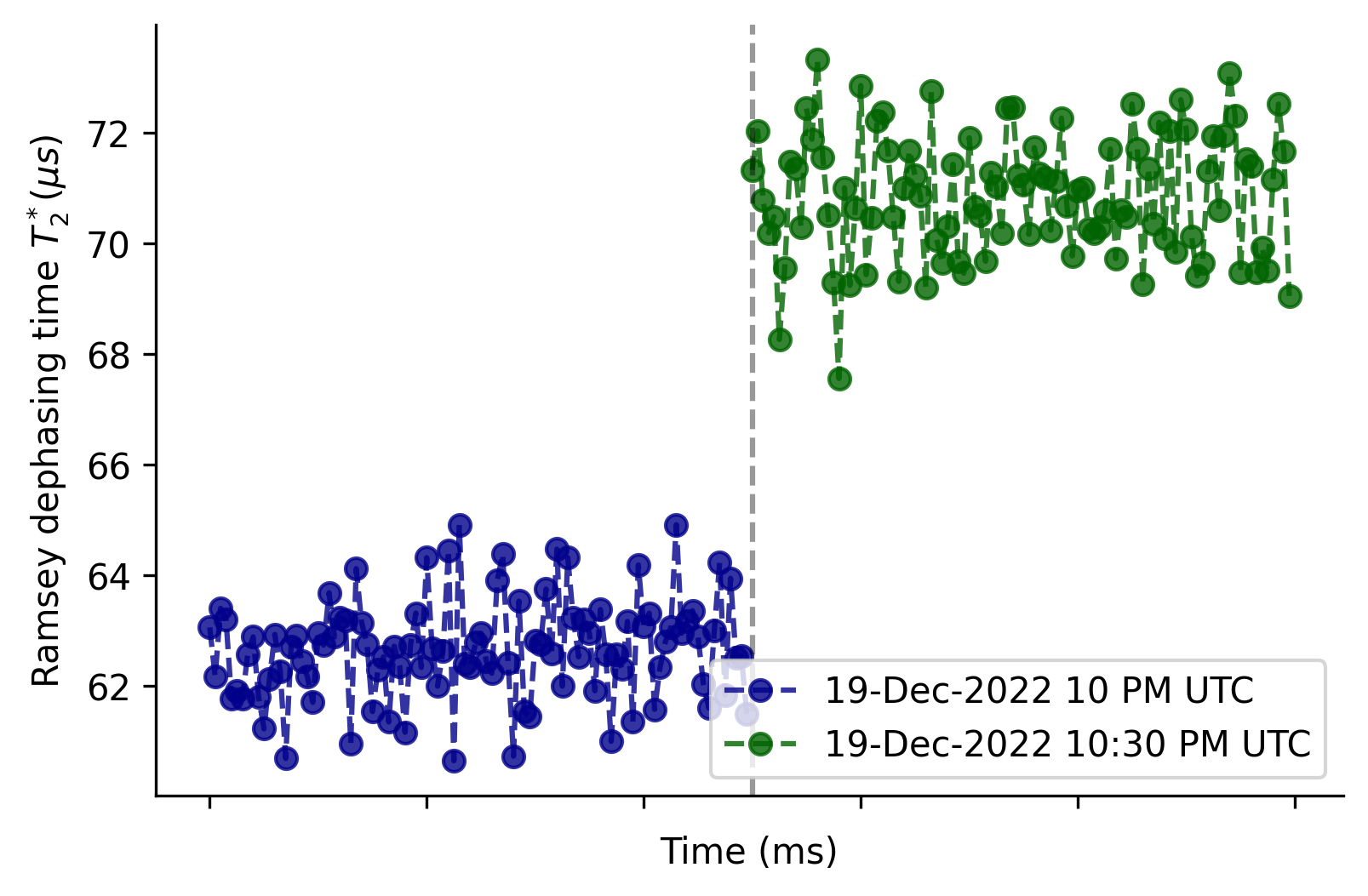}
\caption{This figure presents non-stationary temporal dynamics of decoherence for a qubit on IBM's Belem device. The top figure shows $T_1$ relaxation time series for qubit 0, where two datasets were collected for 5 ms each on the same day, separated by a vertical line. The blue dataset varies between 116-126$\mu$s with a mean of 122 $\mu$s and a standard deviation of 2 $\mu$s, while the green dataset varies between 104-114$\mu$s with a mean of 108 $\mu$s and a standard deviation of 2 $\mu$s. The bottom figure displays Ramsey dephasing time ($T_2$) series for qubit 0, where two datasets were collected for 5 ms each on the same day, separated by a vertical line. The blue dataset varies between 67-73$\mu$s with a mean of 70 $\mu$s and a standard deviation of 1 $\mu$s, while the green dataset varies between 60-64$\mu$s with a mean of 62 $\mu$s and a standard deviation of 1 $\mu$s, collected around different times on the same day. The data shows significant non-stationarity in decoherence values over a 30-minute interval.}
\label{fig:T1T2_temporal}
\end{figure}

\begin{figure}
\centering
\includegraphics[width=.84\columnwidth]{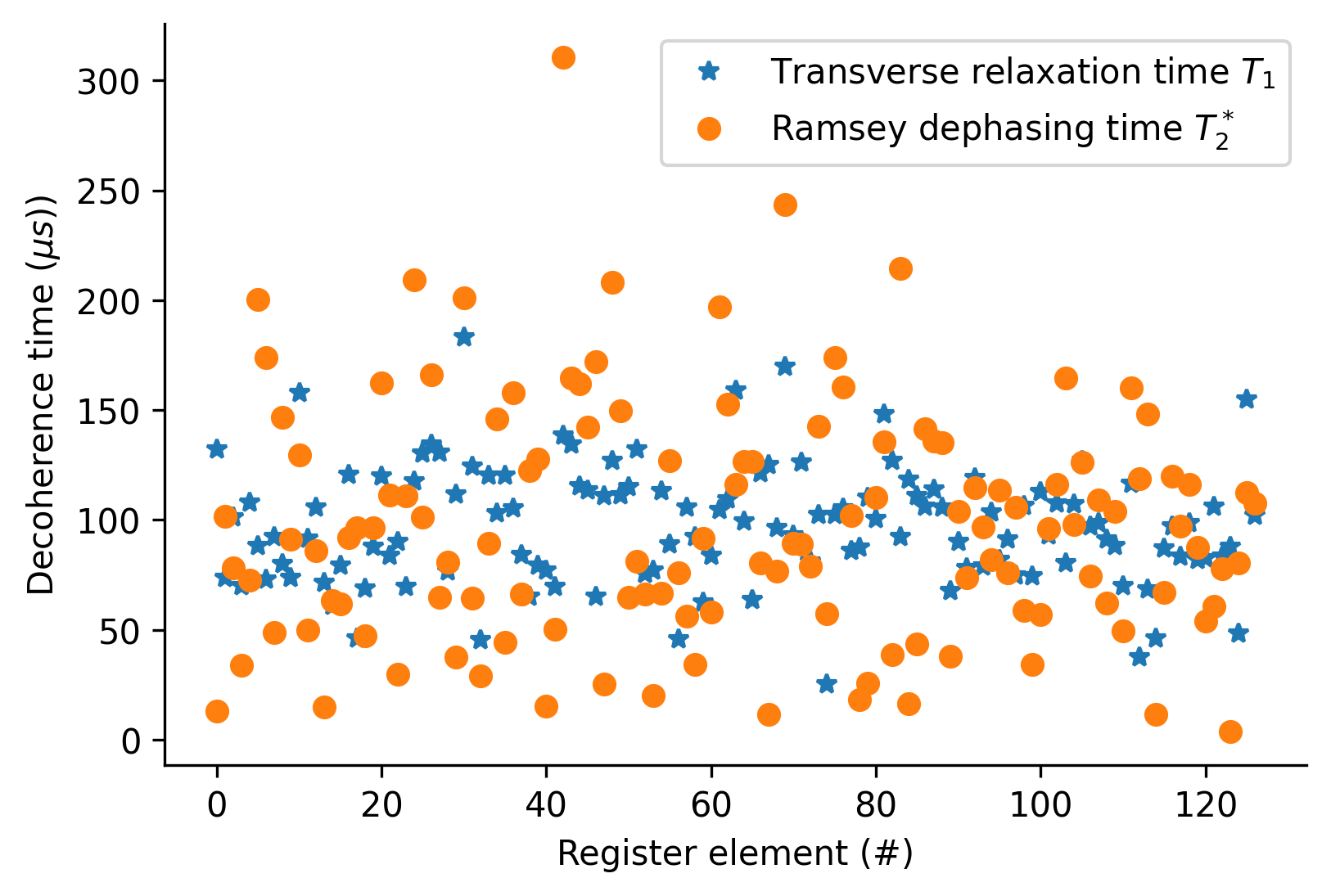}
\caption{This figure displays a snapshot of $T_1$ and $T_2$ times for IBM's 127-qubit transmon device, generated from circuit execution at 14 Jan 2023 10:20 PM UTC. Each of the 127 qubits in the register is represented by a data point for $T_1$ and $T_2$, providing a comprehensive view of the spatial variation of the decoherence times across the register.}
\label{fig:T1T2_spatial}
\end{figure}

Decoherence refers to any loss of unitarity in state evolution, including explicit loss of coherence, energy relaxation effects, and leakage out of the qubit state space. The traditional definition of decoherence, which describes the decay of off-diagonal terms in the density matrix, is now referred to as dephasing and considered one kind of decoherence \cite{hughes2004quantum}. Decoherence studies typically focus on three metrics: transverse relaxation time $(T_1)$, longitudinal relaxation time $(T_2)$, and dephasing time $(T_\phi)$. 
\par 
$T_1$, also known as the transverse relaxation time or relaxation time, measures the attenuation of amplitude in a quantum system. It represents the probability that an excited state $\ket{1}$ will decay to the ground state $\ket{0}$ after time $t$, and is modeled by the function:
\begin{equation}
\textrm{Pr}(\ket{1} \rightarrow \ket{0}) = 1-\exp(-t/T_1) \; .
\end{equation}
The decay-time probability density $f_T(t)$ can be described by the exponential function:
\begin{equation}
f_T(t) = T_1 \exp^{-t/T_1} \; ,
\end{equation}
whose mean is $\mathds{E}(T) = T_1$.
\par 
$T_2$ is a measure of how long it takes for a qubit in the superposition state to decay. Specifically, it measures the decay of the off-diagonal elements of the density matrix and is modeled by an exponential decay function. Therefore, it captures the loss of synchronization between the basis states of an arbitrary quantum ensemble. There are two types of $T_2$ time often quoted in literature \cite{wrightgroup}:
\begin{itemize}
\item Ramsey dephasing time $T_2^*$: measures the time-scale at which a quantum register experiences dephasing effects when left to evolve freely
\item Hahn-echo dephasing time $T_2^{\textrm{echo}}$: uses intermediate $\pi$ pulses for re-focusing to increase relaxation time.
\end{itemize}
When simulating noisy circuits, the appropriate $T_2$ value to use depends on whether the physical implementation of the circuit uses Hahn-echo for noise suppression or not. 
\par 
Finally, the pure dephasing time $(T_\phi)$ is an upper bound on the decoherence time for a qubit, since thermal fluctuations in the environment inevitably cause a loss of phase coherence. However, in practice, the dominant relaxation time is usually $T_2$ (or sometimes $T_1$), rather than $T_\phi$ \cite{wrightgroup}. The three decoherence benchmarks are related by the Bloch-Redfield model:
\begin{equation}
\frac{1}{T_2} = \frac{1}{2T_1} + \frac{1}{T_\phi} \; .
\label{eq:Bloch-Redfield}
\end{equation}
\par 

NISQ devices display random fluctuations in decoherence times $T_1$ and $T_2$. Therefore, the characterization often exhibits significant variability with time and register location. As shown in \cite{carroll2021dynamics, mcrae2021reproducible}, decoherence times can vary significantly from their long-term average, fluctuating by approximately 50\% within an hour. The causes and mechanisms behind decoherence time fluctuations are poorly understood. In transmon registers, potential sources of fluctuations include TLS defects, quasi-particles, parasitic microwave modes, phonons, nuclear spins, paramagnetic impurities, spurious resonances, critical current noise, background charges, gate voltage fluctuations, and the electromagnetic environment \cite{hughes2004quantum}. Among these, TLS defects have been identified as the primary cause of decoherence \cite{bejanin2021interacting, burnett2019decoherence, carroll2021dynamics}. These defects arise from deviations from crystalline order in the naturally occurring oxide layers of transmons, resulting in trapped charges, dangling bonds, tunneling atoms, or collective motion of molecules.
\par 
The fluctuations in $T_1$ and $T_2$ are exemplified by the following findings on IBM's transmon devices. On December 19, 2022, around 9:50 PM UTC, the $T_1$ time of qubit 0 on IBM's Belem device had a narrow variation between 17.04 $\mu$s and 17.33 $\mu$s, with a mean of 17 $\mu$s and a standard deviation of 0.05 $\mu$s. However, the $T_2$ time for the same qubit had a wide variation, ranging from 59 $\mu$s to 65 $\mu$s, with a mean of 62 $\mu$s and a standard deviation of 1 $\mu$s.  This temporal variation at typical program execution time-scales (several hundred micro-seconds) is shown in Fig.~\ref{fig:T1T2_temporal}, which shows the non-stationary temporal dynamics of coherence times in between calibration times.
\par
\begin{figure*}
  \centering
  \begin{tabular}{ c @{\hspace{40pt}} c }
      \includegraphics[width=.8\columnwidth]{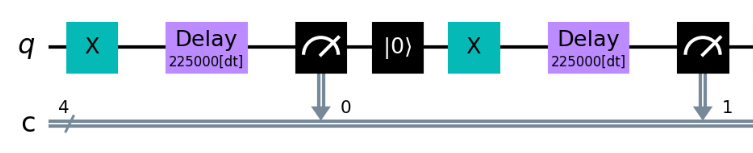} &
    \includegraphics[width=\columnwidth]{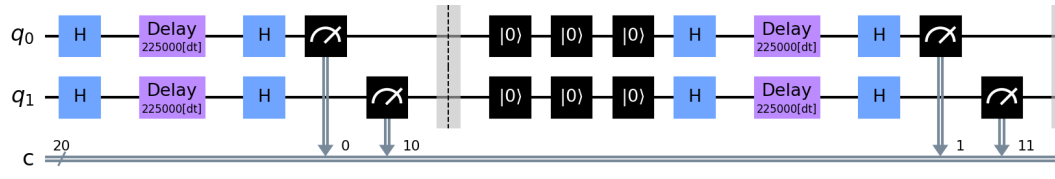} \\
    \small (a) &
      \small (b)
  \end{tabular}
\medskip
\caption{We utilized the depicted circuits to gather (a) $T_1$ and (b) $T_2$ data, on Dec 19, 2022 at 10pm and 10:30pm. Here the delay gates represent a $50 \mu s$ delay (225,000 times the minimum time-interval for such a delay circuit), and qubit reuse is maximized through the software's reset and mid-circuit measurement functionalities, with each estimated data utilizing 100,000 shots.
}
\label{fig:T1T2_subckt}
\end{figure*}

 
When we compare the spatial variation, we find that on January 14, 2023, at 10:20 PM UTC, the $T_1$ time across IBM's Washington device varied widely between 25.3 $\mu$s and 183.1 $\mu$s, with a mean of 97.2 $\mu$s and a standard deviation of 26.8 $\mu$s. On the same date and time, the $T_2$ time for the qubits on the same device showed an even broader variation, ranging from 3.7 $\mu$s to 310.4 $\mu$s, with a mean of 97.5 $\mu$s and a standard deviation of 54.1 $\mu$s. This is shown in Fig.~\ref{fig:T1T2_spatial}. Note that data on $T_2$ times for all qubits of the Washington register, after calibration on a specific day, is publicly accessible through IBM data servers and can be obtained using Qiskit. The circuits utilized by us to gather $T_1$ and $T_2$ data are depicted in Fig.~\ref{fig:T1T2_subckt}~(a) and (b) respectively. In these circuits, the delay gates represent a $50 \mu s$ delay, equivalent to 225,000 times the minimum time-interval achievable for such a delay circuit. To maximize qubit reuse, the software's reset and mid-circuit measurement functionalities are employed. Each estimated data used 100,000 shots.

\par 
The decoherence time fluctuations in both temporal and spatial domains can lead to inaccurate error mitigation/correction, when verifying test circuits, and unreliable quantum program output in general. Therefore, it is crucial to quantify the uncertainty of these characterizations and identify their sources and mechanisms. This will allow researchers to strike a balance between reporting reliability and precision.

\section{Spatio-temporal stochasticity of the Pauli noise channel} \label{sec:model}
\begin{figure}[!h]
\centering
\includegraphics[width=.84\columnwidth]{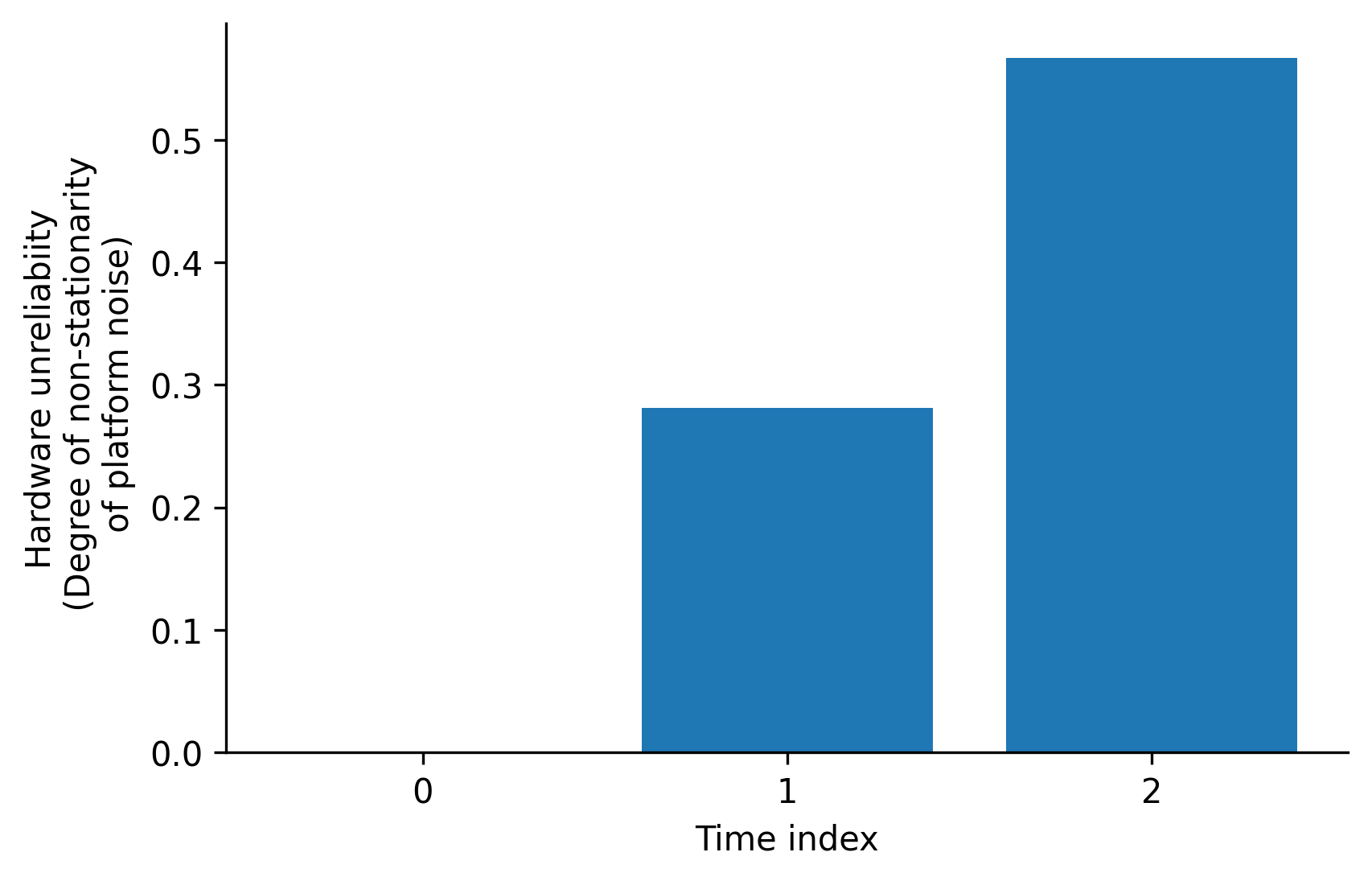}
\caption{We model the degradation of a non-stationary Pauli noise channel by assuming the coherence time steadily decreases over time. Using a non-stationary Dirichlet distribution, we model the joint distribution of coefficients for a two-qubit circuit, which fluctuate as coherence times deteriorate. The y-axis represents the degree of non-stationarity, and the x-axis shows four time-periods. The Hellinger distance between the Dirichlet distributions at time $\tau$=0 and a later time is a measure of non-stationarity, increasing from 0 to 57\%. This model is based on transmon platforms and is used as an experimental setup for our simulation experiments.}
\label{fig:channel_quality}
\end{figure}
\subsection{Single-qubit channel}
The amplitude damping channel $\mathcal{E}^\textrm{AD}(\cdot)$ and dephasing channel $\mathcal{E}^\textrm{PD}(\cdot)$ are two fundamental sources of quantum de-coherence and information loss in transmons \cite{burnett2019decoherence, klimov2018fluctuations, nielsen2002quantum}. A realistic model for this noise channel, denoted as APD, involves a combination of amplitude damping and dephasing.
\par 
Amplitude damping can be described by the Kraus operators $E_0^{\textrm{AD}}$ and $E_1^{\textrm{AD}}$, while phase damping can be described by $E_0^{\textrm{PD}}$ and $E_1^{\textrm{PD}}$, as follows \cite{nielsen2002quantum}:
\begin{equation}
\mathcal{E}^{\textrm{AD}}(\rho) = \sum\limits_{k=0}^1 E_k^{\textrm{AD}}\rho E_k^{\textrm{AD}\dagger} \; ,
\end{equation}
\begin{equation}
E_0^{\textrm{AD}} = 
\begin{pmatrix}
1 & 0\\
0 & \sqrt{1-\gamma}
\end{pmatrix} \; ,
\end{equation}
\begin{equation}
E_1^{\textrm{AD}} = 
\begin{pmatrix}
0 & \sqrt{\gamma}\\
0 & 0
\end{pmatrix} \; ,
\end{equation}
\begin{equation}
\mathcal{E}^{\textrm{PD}}(\rho) = \sum\limits_{k=0}^1 E_k^{\textrm{PD}} \rho E_k^{\textrm{PD}\dagger} \; ,
\end{equation}
\begin{equation}
E_0^{\textrm{PD}} = 
\begin{pmatrix}
1 & 0\\
0 & \sqrt{1-\lambda}
\end{pmatrix} \; ,
\end{equation}
\begin{equation}
E_1^{\textrm{PD}} = 
\begin{pmatrix}
0 & 0\\
0 & \sqrt{\lambda}
\end{pmatrix} \; .
\end{equation}
Here, $\gamma = 1-\exp(-t/T_1)$ and $\lambda = 1-\exp(-t/T_2)$, where $t$ is the time scale of the decoherence process. The relation between $T_1$, $T_\phi$, and $T_2$ was previously discussed in Eq.~(\ref{eq:Bloch-Redfield}). The Kraus decomposition of the combined amplitude and phase damping channel $\mathcal{E}^\textrm{APD}(\cdot)$, valid for a single qubit, can be expressed as $E_0^{\textrm{APD}}$, $E_1^{\textrm{APD}}$, and $E_2^{\textrm{APD}}$. 
\begin{equation}
\mathcal{E}^{\textrm{APD}}(\rho) \coloneqq \mathcal{E}^{\textrm{PD}}\circ \mathcal{E}^{\textrm{AD}} = \sum\limits_{k=0}^2 E_k^{\textrm{APD}} \rho E_k^{\textrm{APD}^\dagger} \; ,
\end{equation}
where, 
\begin{equation}
E_0^{\textrm{APD}} = E_0^{\textrm{PD}} E_0^{\textrm{AD}} =
\begin{pmatrix}
1 & 0\\
0 & \sqrt{[1-\gamma][1-\lambda]}
\end{pmatrix} \; ,
\end{equation}
\begin{equation}
E_1^{\textrm{APD}} = E_0^{\textrm{PD}} E_1^{\textrm{AD}} = 
\begin{pmatrix}
0 & \sqrt{\gamma}\\
0 & 0
\end{pmatrix} \; ,
\end{equation}
\begin{equation}
E_2^{\textrm{APD}} = E_1^{\textrm{PD}} E_0^{\textrm{AD}} =
\begin{pmatrix}
0 & 0\\
0 & \sqrt{[1-\gamma]\lambda}
\end{pmatrix} \; .
\end{equation}
Using the fact that:
\begin{align}
    & E_0^{\textrm{APD}} = \frac{1+\sqrt{1-\lambda-\gamma+\lambda\gamma}}{2}I + \frac{1-\sqrt{1-\lambda-\gamma+\lambda\gamma}}{2}Z \; , \\
    & E_1^{\textrm{APD}} = \frac{\sqrt{\gamma}}{2}X+  \frac{\sqrt{\gamma}}{2}iY \; , \\
    & E_2^{\textrm{APD}} = \frac{\sqrt{\lambda-\lambda\gamma}}{2}I - \frac{\sqrt{\lambda-\lambda\gamma}}{2}Z \; .
\end{align}
the APD channel can be expressed as:
\begin{equation}
\begin{split}
\mathcal{E}^{\textrm{APD}}(\rho) =& \frac{2-\gamma+2\sqrt{1-\lambda-\gamma+\lambda\gamma}}{4}\rho
+\frac{\gamma}{4} X\rho X
-\frac{\gamma}{4} Y\rho Y\\
&+\frac{2-\gamma-2\sqrt{1-\lambda-\gamma+\lambda\gamma}}{4} Z \rho Z\\
&+\frac{\gamma}{4} I \rho Z
+\frac{\gamma}{4} Z \rho I
-\frac{\gamma}{4i} X \rho Y
-\frac{\gamma}{4i} Y \rho X \; ,
\end{split}
\end{equation}
where $\lambda$ and $\gamma$ are the APD parameters, and $I$, $X$, $Y$, and $Z$ are the Pauli matrices.
\par 
Multi-qubit and multi-gate APD channels are not efficiently simulatable due to the Gottesman-Knill theorem \cite{gottesman1998heisenberg}, which states that only Pauli channels, those with Pauli matrices as their Kraus operators, can be simulated efficiently on a classical computer. Channel twirling maps \cite{eggeling2001separability, dankert2009exact, magesan2008gaining} a complex channel to a simpler one while preserving certain features such as the average channel fidelity and the entanglement fidelity \cite{horodecki1999general}. In particular, it transforms a non-Pauli channel into the Pauli noise channel, which is widely used in quantum error correction because it is a simple and natural model for random quantum noise \cite{emerson2007symmetrized, silva2008scalable, martinez2020approximating}. It is a well-understood and easily implementable noise model that can simulate a variety of realistic physical processes that lead to quantum errors, such as dephasing, amplitude damping, and phase-flip errors. Additionally, the Pauli noise channel is mathematically tractable and can be efficiently simulated, making it a useful tool for developing and testing quantum error correction protocols. It can be used to estimate the average fidelity of a quantum gate subject to the original APD channel and identify codes that work for the APD channel \cite{silva2008scalable}.
\par 
Thus, using the fact that the average channel fidelities of the APD and its twirled version (i.e., the Pauli noise channel) are the same, one can express the Pauli noise parameters as a function of the $T_1$ and $T_2$ times of the APD. 
\par 
Pauli twirling is defined as:
\begin{equation}
\mathcal{E}_\textrm{twirl}(\rho) = \frac{1}{4} \sum\limits_{A \in \{I, X, Y, Z\}} A^\dagger \mathcal{E}\left( A\rho A^\dagger \right)A \; .
\end{equation}
This is a convex combination of channels and hence also a channel. 
Specifically, The twirled APD is a Pauli channel and hence efficiently simulable. Its noise parameters, expressed as a function of the $T_1$ and $T_2$ times of the APD, can be used to model the noise with the APD channel while also benefiting from the simplicity of the Pauli noise channel for efficient simulations. The expression for twirled APD is:
\begin{equation}
\begin{split}
\mathcal{E}_\textrm{twirl}^{\textrm{APD}} \left( \rho; T_1, T_2 \right)  =& 
\frac{1}{4} \sum\limits_{A \in \{I, X, Y, Z\}} A^\dagger \mathcal{E}^{\textrm{APD}}\left( A\rho A^\dagger \right)A\\
=& \sum\limits_{k=0}^3 c_k \sigma_k \rho \sigma_k \; ,
\end{split}
\end{equation}
where,
\begin{equation}
\begin{split}
c_1 = c_2 = & \frac{1}{4}\left[1-\exp\left(-t/T_1\right)\right] \; , \\
c_3 =& \frac{1}{4}\left[1-\exp\left(-t/T_2\right)\right] \; , \\
c_0 =& 1- (c_1 + c_2 + c_3) \; , \\
\end{split}
\label{eq:depol_coeffs}
\end{equation}
and $\{ \sigma_k \}_{k=0}^3 = \{I, X, Y, Z\}$ are the Pauli matrices. Note that the Pauli error probabilities $c_1, c_2, c_3$ are not equal to each other. 
Section~\ref{sec:decoherence} discussed the random fluctuations of the decoherence times $T_1$ and $T_2$ as spatially and temporally varying stochastic processes across all time-scales. We express these fluctuations as random variables dependent on register location $i$ and time $\tau$:
\begin{equation}
T_1 = T_1(i,\tau) \quad \text{and} \quad T_2 = T_2(i,\tau) \; .
\label{eq:T_i_t}
\end{equation}
It follows from Eq.~(\ref{eq:depol_coeffs}) that the coefficients of the single-qubit Pauli noise channel will fluctuate randomly as temporally and spatially varying stochastic processes \cite{etxezarreta2021time}. These random variables also depend on register location $i$ and time $\tau$:
\begin{equation}
c_0 = c_0(i,\tau), \;\; c_1 = c_1(i,\tau), \;\; c_2 = c_2(i,\tau), \;\; c_3 = c_3(i,\tau) \; .
\label{eq:x_i_t}
\end{equation}

\subsection{Multi-qubit channel}\label{sec:TSVdepol}
\begin{table}[htbp]
\caption{True means of the time-varying noise Pauli channel coefficients}
\begin{center}
\begin{tabular}{c c c c c}
\textrm{Pauli term [$\textrm{qubit} 0 \otimes \textrm{qubit} 1$]} & \textrm{Period 0} & \textrm{Period 1} & \textrm{Period 2}&\\
\toprule
II&0.379&0.326&0.26&\\
IX&0.056&0.064&0.075&\\
IY&0.056&0.064&0.075&\\
IZ&0.076&0.078&0.079&\\
XI&0.081&0.083&0.082&\\
XX&0.012&0.016&0.024&\\
XY&0.012&0.016&0.024&\\
XZ&0.016&0.02&0.025&\\
YI&0.081&0.083&0.082&\\
YX&0.012&0.016&0.024&\\
YY&0.012&0.016&0.024&\\
YZ&0.016&0.02&0.025&\\
ZI&0.127&0.12&0.108&\\
ZX&0.019&0.024&0.031&\\
ZY&0.019&0.024&0.031&\\
ZZ&0.026&0.029&0.033&\\
\bottomrule
\end{tabular}
\end{center}
\label{tab:coefficients_true}
\end{table}
Let us characterize the effect of Pauli noise on quantum information encoded in an n-qubit register:
\begin{equation}
\mathcal{E}_\xrm(\rho) = \sum\limits_{i=0}^{N_p-1} \xrm_i P_i(n) \rho P_i(n)^\dagger \; ,
\end{equation}
where $N_p=4^n$ is the total number of Pauli coefficients and $P_i(n)$ are the n-qubit Pauli operators with coefficients subject to the conditions:
\begin{equation}
\sum\limits_{i=0}^{N_p-1}\xrm_i=1, \;\;\;\; \xrm_i \geq 0 \; .
\label{eq:simplex}
\end{equation}
To study the performance of error mitigation and correction algorithms in non-stationary NISQ devices, it is important to investigate the behavior of a non-stationary Pauli noise channel. The $N_p$ coefficients of the Pauli operators can be considered as random variables that fluctuate spatially and temporally, denoted as $\xrm_k = \xrm_k(i, \tau)$, where $i = (i_1, \cdots, i_n)$ identifies the register location(s) and $\tau$ is time. A prior hypothesis for the Pauli channel  can be obtained by assuming channel separability, in which case the $N_p$ coefficients can be obtained using a direct product over $n$ four-dimensional vectors, as shown in Eq.~(\ref{eq:direct_product}). 
\begin{equation}
\begin{split}
\xrm &= 
\begin{pmatrix}
c_0(i=0,\tau)\\
c_1(i=0,\tau)\\
c_2(i=0,\tau)\\
c_3(i=0,\tau)\\
\end{pmatrix}
\times \cdots \times 
\begin{pmatrix}
c_0(i=n-1,\tau)\\
c_1(i=n-1,\tau)\\
c_2(i=n-1,\tau)\\
c_3(i=n-1,\tau)\\
\end{pmatrix} \; ,
\label{eq:direct_product}
\end{split}
\end{equation}
where $\times$ refers to the direct product. 
\par 
To incorporate correlations between qubit errors, one could still consider a separable noise model (for the simplicity of modeling the noise using single-qubit APD channels). These correlations stem from the spatial correlation function (for a fixed time $\tau$ and error type $k$) between qubits $i=a$ and $i=b$ on the device
\begin{equation}
    \overline{\xrm_{k}(i=a,\tau)\xrm_{k}(i=b,\tau)} \; , \label{eq:space_corr}
\end{equation}
and fitting it to a pre-determined model, e.g. an exponentially decaying function (the decay constant quantifies how strongly the different qubit noises are correlated). Here, the average is taken over the realizations of the circuit. Now, since we do not really measure $\xrm_{k}$ in our circuits, instead we measure $T_{1}(i=a, \tau)$ and $T_{2}(i=a, \tau)$, we can substitute for $x_{k}(i=a, \tau)$ its dependence on $T_{1}$ and $T_{2}$, given the twirling of a specific noise model (this is done analytically above for the single-qubit APD channel), and then the spatial and temporal statistical properties of $T_{1}$ and $T_{2}$ over our sample could be used to compute these qubit-qubit noise correlations. This may provide a richer characterization of the n-qubit case that goes a step beyond the standard ``independent'' noise approximation, which seems to fail for the particular IBM quantum computer (and likely all quantum computers more generally).
\par 
Note that in this analysis, there are two layers to the ``independent'' noise model that one can consider. (1) we can say that any Pauli channel satisfying Eq.~\eqref{eq:direct_product} is an independent noise model. However, if Eq.~\eqref{eq:space_corr} is non-zero for at least two qubits $a \neq b$, then the Pauli errors still contain classical (non-entangling) spatial correlations between them. (2) A subset of Pauli channels satisfying Eq.~\eqref{eq:direct_product} will also nullify Eq.~\eqref{eq:space_corr}. 
%
\subsection{Quantifying the non-stationarity}
The Pauli noise channel, although not a completely general noise model, still manages to model many practical situations. It is widely used because of two reasons: (a) it is efficiently simulatable on a classical computer (per the Gottesman-Knill theorem) and (b) when used as a proxy for physically accurate noise models (such as the amplitude and phase damping noise) which are not efficiently simulatable on a classical computer, it still manages to preserve important properties like entanglement fidelity \cite{horodecki1999general}.
\par 
As discussed in Section 2, the coefficients of the Pauli noise channel (henceforth referred to as the Pauli coefficients), are directly linked to the decoherence times $T_1$ and $T_2$ of the individual register elements.
\par 
These decoherence times exhibit stochastic fluctuations in time and across the register and show correlations between them. As a result, the Pauli coefficients also show strong correlations among them. Another constraint is given by Eq.~\eqref{eq:direct_product}. Because $\sum \xrm_i = 1$ and each $\xrm_i \in [0,1]$, the  natural way to model the probability distribution function  $f_X(\xrm)$ of the multi-dimensional Pauli channel distribution is the Dirichlet distribution:
\begin{equation}
f_X(\xrm) \equiv \textrm{Dirichlet}(\xrm; \eta)  \coloneqq  
\frac{ \Gamma \left( \sum\limits_{i=0}^{N_p-1} \eta_i \right)}
{\prod\limits_{i=0}^{N_p-1}\Gamma(\eta_i)}\left(\prod\limits_{i=0}^{N_p-1}\xrm_i^{\eta_i-1}\right) \; ,
\end{equation}
where $\eta \ge 0$ is the Dirichlet hyper-parameter for the Pauli channel distribution, $\Gamma$ is the Gamma function, and $\int\limits_{\xrm} \textrm{Dirichlet(x; $\eta$) dx} = 1$.
\par As discussed in the previous two sections, the distribution of the decoherence times fluctuates with time. Thus, the distribution of the Pauli coefficients (i.e. the Dirichlet distribution), changes with time. Specifically, the coefficients $\eta_i$ will vary with time. The distribution may be represented as $f_X(\xrm; \tau)$. At any time instant (or at any circuit execution instance), the Pauli coefficient vector x represents a specific realization of the random variable X, which is sampled from $f_X(\xrm; \tau)$.
\par 
Recall that we measure reliability using the Hellinger distance $H_d$ between the distribution of the parameters characterizing the noise (in this case, the Pauli coefficients). The Hellinger distance between two time-varying Dirichlet distributions (parameterized by $\eta$ and $\eta^\prime$), quantifies the degree of non-stationarity:
\begin{equation}
\begin{aligned}
& H_d \coloneqq \sqrt{1-B C} \; , \\
& BC \coloneqq \int\limits_\xrm \sqrt{f(\xrm) g(\xrm)} d\xrm\\
&=\frac{
\sqrt{
\Gamma\left( \sum\limits_{i=0}^{N_p-1} \eta_{i} \right)
\Gamma\left( \sum\limits_{i=0}^{N_p-1} \eta_{i}^{\prime}\right)
}}
{\prod\limits_{i=0}^{N_p-1} 
\sqrt{
\Gamma(\eta_{i}) \Gamma(\eta_{i}^{\prime})
}}\times \frac{
\prod\limits_{i=0}^{N_p-1}
\Gamma\left(\frac{\eta_{i}+\eta_{i}^{\prime}}{2}\right)}
{\Gamma\left(\frac{ \sum\limits_{i=0}^{N_p-1}  \eta_{i}+\eta_{i}^{\prime}}{2}\right)} \; ,
\end{aligned}
\end{equation}
where $N_p=4^{n}$ and BC is the Bhattacharya coefficient.
\section{Bayesian inference of stochastic Pauli noise}\label{sec:bayesian}
Bayesian statistical models treat unknowns as random variables and require a prior, while frequentist statistical models treat unknowns as parameters and rely solely on a data generating model. Bayesian estimators rely on the posterior for inference. Adaptive estimation using Bayesian inference can improve the performance of error mitigation schemes in the presence of time-varying noise. This approach updates the system noise hypothesis as new information becomes available. Bayesian updating is particularly important for Probabilistic Error Cancellation (discussed in the next section) when dealing with time-varying noise. 
\par 
In particular, the time-varying Dirichlet distribution of the Pauli coefficient should be dynamically estimated to be able to mitigate errors more accurately. Dynamic estimation can be done either through a point-in-time update, which uses maximum likelihood estimation, or a Bayesian inference-based rolling update. The Bayesian method is especially useful when dealing with sparse data. Bayesian inference of the Pauli noise channel is simply an application of Bayes' rule:
\begin{equation}
\begin{aligned}
f_X(\xrm | \textrm{observed data})
&\propto f_X(\textrm{observed data} | \xrm )  f^\textrm{prior}_X(\xrm) \; .
\end{aligned}
\end{equation}

The prior distribution for the Pauli channel coefficients is given by:
\begin{equation}
\begin{aligned}
\xrm &\sim f^\textrm{prior}_X(\xrm) = \textrm{Dirichlet}(\xrm; \eta)\\
&= \frac{ \Gamma \left( \sum\limits_{j=0}^{N_p-1} \eta_j \right)}
{\prod\limits_{j=0}^{N_p-1}\Gamma(\eta_j)}\left(\prod\limits_{j=0}^{N_p-1}\xrm_i^{\eta_j-1}\right) \; .
\end{aligned}
\end{equation}
We generate our data set by executing the quantum circuit L times and recording the outcome s of the circuit measurement POVM $\{\Pi_{s}\}$, where s is an integer representing the observed n-bit long binary string. The data set is represented by the set $\{s_l\}$ where $l \in 0, \cdots, L-1$. Each execution of the circuit results in a specific outcome s. The random variable S follows a discrete distribution from the range of $0, \cdots, N-1$ with probabilities $p_0, p_1, \cdots,  p_{N-1}$,
where $N=2^n$ is the Hilbert space dimension. Also:
\begin{equation}
p_s = \textrm{Tr}[\Pi_s \tilde{\mathcal{G}}_{\hat{\xrm}}(\rho_\textrm{test})] \; ,
\end{equation}
where $\tilde{\mathcal{G}}_\xrm$ is a noisy implementation of the ideal quantum operation $\mathcal{G}$ characterized by the noise parameter x and $\rho_\textrm{test}$ is a known test density matrix.
\par 
Assuming circuit runs are independent of each other, the likelihood function is given by:
\begin{equation}
\begin{aligned}
\Pr( \{ s_\ell \} | x) =
&=\prod_{i=0}^{N-1} \left[ \textrm{Tr}[\Pi_i \tilde{\mathcal{G}}_\xrm(\rho_\textrm{test})] \right]^{C_i(\{ s_\ell \})} \; , \\
\end{aligned}
\end{equation}
where, $C_i(\{ s_\ell \}) = \sum\limits_{\ell=0}^{L-1} \delta_i(s_\ell)$ (with $i \in \{0, 1, \cdots N-1\}$) is simply a counter function that counts how many times $i$ appeared in the experimentally observed data post-measurement ($\delta_i(s)$ is Kronecker delta function which is 1 if $i=s$ and zero otherwise). The log of the posterior is then given by:
\begin{equation}
\begin{aligned}
\log f^\textrm{posterior}_X(\xrm | \textrm{data}) 
&= \sum\limits_{i=0}^{N-1} C_i(\textrm{data}) \log \left[ \textrm{Tr}[\Pi_i \tilde{\mathcal{G}}_\xrm(\rho_\textrm{test})] \right]\\
&+\sum\limits_{j=0}^{N_p-1}(\eta_j-1)\log \xrm_j\\
&+ \textrm{terms independent of x} \; .
\end{aligned}
\end{equation}
Lastly, the Maximum-a-posterior (MAP) estimate is obtained as:
\begin{equation}
\begin{aligned}
\hat{\xrm}(\tau) =& \underset{\xrm}{\textrm{argmax}} \left[ \sum\limits_{i=0}^{N-1} C_i(\textrm{data}) \log \left[ \textrm{Tr}[\Pi_i \tilde{\mathcal{G}}_\xrm(\rho_\textrm{test})] \right]\right.\\
&\left. +\sum\limits_{j=0}^{N_p-1}(\eta_j-1)\log \xrm_j\right] \; .
\end{aligned}
\end{equation}
\subsection{Example application: PEC on unreliable platforms}
\subsubsection{Method}
Probabilistic error cancellation (PEC) is a well-known error mitigation method \cite{temme2017error, endo2018practical, zhang2020error}. The four broad steps of the PEC workflow are as follows. We use the convention that calligraphic symbols denote super-operators acting on density matrices: 
\begin{equation}
\mathcal{G}(\rho)=G\rho G^\dagger.
\end{equation}
\par 
First, expand an ideal unitary gate $\mathcal{G}$ as a (noise-model dependent) linear combination of implementable noisy gate set $\{\tilde{\mathcal{G}}_j\}$ (with its ideal counterpart $\{\mathcal{G}_j\}$), as follows:
\begin{equation}
\mathcal{G} = \sum\limits_{j=0}^{N_p-1} \theta_j \tilde{\mathcal{G}}_j,
\end{equation}
where $\theta_j$ are real coefficients, and $\mathcal{E}_\xrm$ is an error channel (such as Pauli noise channel), and $\tilde{\mathcal{G}_j}\equiv \mathcal{E}_\xrm \circ \mathcal{G}_\textrm{j}$. For example, if $G=H\otimes H$ is a two-qubit Hadamard gate, then the noisy basis circuits are given by 
\begin{equation}
\tilde{\mathcal{G}}_{\sigma \sigma^{\prime}} = \mathcal{E}_\xrm \circ \mathcal{P}_{\sigma \sigma^{\prime}} \circ \mathcal{G}
\end{equation}
where $\mathcal{P}_{\sigma \sigma^{\prime}}(\cdot)\equiv (\sigma \otimes \sigma^{\prime})(\cdot)(\sigma \otimes \sigma^{\prime})$, and $\sigma, \sigma^{\prime}$ are picked from the set of Pauli matrices $\{I,X,Y,Z\}$.
\par 
The second step of PEC involves using a measurement-based estimation procedure to estimate the expectation value of the mitigated operator as $\sum\limits_{j=0}^{N_p-1} \theta_j \braket{\tilde{\mathcal{G}}_j}$. 
To improve computational efficiency, we sample an implementable gate from the quasi-probability representation of the ideal gate, using a probabilistic approximation with Monte Carlo importance sampling. The ideal gate can be approximated as:
\begin{equation}
\mathcal{G} = \sum\limits_{w=0}^{N_p-1} p(w) [\Theta \textrm{sgn}(\theta_w) \tilde{\mathcal{G}}_w],
\end{equation}
where 
\begin{equation}
\Theta=\sum\limits_{w=0}^{N_p-1} |\theta_w|
\end{equation}
and $p(w) = |\theta_w|/\Theta$ and $w$ is a random variable such that $w \in {0,1,\cdots,N_p-1}$. The procedure gives more weight to coefficients $\theta_w$ with larger magnitudes. The key fact behind PEC is that, given the random variable $\tilde{w}$, the probabilistic super-operator $\Theta \textrm{sgn}(\theta_w) \tilde{\mathcal{G}}_w$ is an unbiased estimator for the ideal gate $\mathcal{G}$, i.e. $\mathcal{G}=
\mathds{E}_w \left[ \Theta \textrm{sgn}(\theta_w) \tilde{\mathcal{G}}_w\right]$.
\par 
The third step of PEC involves sampling an implementable circuit from the quasi-probability representation of an ideal circuit. This is achieved by sampling from the quasi-probability representation of each ideal gate of the circuit, multiplying the sign of each sampled gate to obtain the global sign associated with the full circuit, and multiplying the $\Theta$ of each sampled gate to obtain the global $\Theta$ associated with the full circuit. This produces an unbiased estimator of the ideal circuit, and the sampling average of the circuit estimator converges (in the limit of many samples) to the ideal circuit, similar to the previous case of a single-gate estimator.
\par 
The final step of PEC involves inferring an ideal expectation value from the noisy execution of the sampled circuits. In our 2-qubit example application, we take the observables as the projector on the computation basis states to infer the ideal histogram. There are four projectors, namely $\Pi_0=\ket{00}\bra{00}$, $\Pi_1=\ket{01}\bra{01}$, $\Pi_2=\ket{10}\bra{10}$, and $\Pi_3=\ket{11}\bra{11}$, each representing a computation basis state. To obtain a mitigated estimate of the ideal probability associated with a basis state, one needs to sample many circuits from the quasi-probability representation of the ideal circuit, execute all the samples with the noisy executor, and obtain a list of noisy expectation values for each projector. Then, by averaging with suitable weights (signs and $\Theta$), the ideal histogram can be obtained.
\par 
We immediately note that PEC provides the ideal circuit in the asymptotic Monte Carlo circuit limit if the noise is static and perfectly characterized, which does not require updating the PEC implementation. However, in the presence of non-stationary time-varying noise distributions, an adaptive approach becomes necessary.
\subsubsection{Simulation setup}
The time-variation in the Pauli noise channel on the transmon platform highlights its unreliability. This poses challenges for any algorithm that relies on accurate noise characterization, including probabilistic error cancellation (PEC), which suffers from such reliability issues.
\par 
PEC is a method for approximating the noiseless ideal sub-system of an unknown circuit made up of smaller modular sub-systems such as CNOT, X, Y, Z, and Hadamard gates. By generating and averaging many Monte Carlo circuits, we can construct a "noiseless ideal sub-system" which can be used to implement a bigger circuit. PEC assumes an effective noise model that affects basis circuits. These noisy basis circuits are assumed to be implemented perfectly, as detailed in \cite{temme2017error, larose2022mitiq}. Therefore, accurately characterizing the noise is crucial for successful PEC implementation. 
The time-varying noise in the platform makes the basis circuit set used by PEC itself time-varying, requiring adaptive channel estimation each time.
\par 
In this section, our goal is to implement a noiseless, ideal operation on two qubits in the presence of time-varying noise using adaptive estimation and probabilistic error cancellation. Specifically, we aim to execute a Hadamard operation on qubit 0 and qubit 1, denoted as $H\otimes H$, when only noisy basis gates are available. This is achieved in the presence of general, non-separable, time-varying Pauli noise channel. 
\subsubsection{Time-varying Pauli noise}
In our simulation, we consider a two-qubit register subject to time-varying noise. The experimental data and causes of the decoherence fluctuations are discussed in Section~\ref{sec:decoherence}. The data used in our experiment was obtained from a transmon device called Belem by IBM in December 19, 2022.
\par 
We assume that the mean of the stochastic $T_1$ coherence time for the first qubit decreases uniformly in a simple step-function-like manner over five time periods, deteriorating from 150 to 60 $\mu$s. Similarly, for the second qubit, we assume that the mean of the stochastic $T_1$ coherence time deteriorates from 200 to 10 $\mu$s in five time periods. Additionally, we assume a simple step-function-like decrease in the mean of the stochastic $T_2$ coherence time for the first qubit, deteriorating from 70 to 50 $\mu$s in five time periods. Finally, for the second qubit, we assume that the mean of the stochastic $T_2$ coherence time will deteriorate from 130 to 62.5 $\mu$s in five time periods. The coefficients of the Pauli channel are then computed using Eq.~(\ref{eq:depol_coeffs}). This setup models the performance when the non-identity Pauli channel coefficients increases with time, such as in between calibrations. We assume a typical execution time of 100 $\mu$s for the Hadamard gate on the IBM transmon platform.
\par 
We investigate the non-stationarity of the time-varying Pauli noise channel in our simulation where the register coherence time decreases steadily with time. We model the joint distribution of the coefficients using a non-stationary Dirichlet distribution, which varies with time. The Hellinger distance between the Dirichlet distribution at time $\tau$=0 and a later time is a measure of the degree of non-stationarity, increasing from 0 to 57\% as shown in Fig.~\ref{fig:channel_quality}. The true means of the time-varying noise Pauli channel coefficients are shown in Table~\ref{tab:coefficients_true}.
\par 
In period 0, the coefficient of the identity term in the Pauli noise channel is 38\%, but by period 2, it degrades to 26\%. This change is driven by the deterioration in the coherence times for qubit 0 and qubit 1, respectively, as described above. We use the Hellinger distance to maintain consistency with the Bayesian inference section as the Dirichlet distribution will be used later in this study.
\subsubsection{Initial time period}
At the outset, we assume we are in period 0, equipped with accurate knowledge of the Pauli noise channel. With this knowledge, we obtain the super-operator expression for the noisy basis circuits, which we linearly combine to estimate the ideal operation ($H \otimes H$). This linear combination is almost error-free, as long as the noisy basis circuits expressed as super-operators are completely positive and trace preserving (CPTP). Post importance sampling using the PEC procedure, we obtain almost error-free results for the mean of the observable. The reconstructed operation is a weighted average using a quasi-probability distribution that uses the true noisy basis. Our results, shown in Fig.~\ref{fig:output_quality}~(b), indicate an accurate ideal gate implementation in the presence of noise in period 0, with a Hellinger distance between the expected and observed output of 0.34\%. This small error stems from the finite Markov chain Monte Carlo (MCMC) circuit samples.
\subsubsection{Subsequent time periods}
In subsequent time periods, the noise characteristics of the Pauli channel change as shown in Fig.~\ref{fig:channel_quality} and Table~\ref{tab:coefficients_true}. This change renders the previously implemented PEC approach inaccurate as the super-operators characterizing the noisy basis circuits are no longer accurate. Without adaptive mitigation, two issues arise: the old linear combination coefficients for importance sampling of the Monte-Carlo circuits are no longer valid, and the circuits, post generation, are impacted by new types of noise during execution. 
The black bars in Fig.~\ref{fig:output_quality}~(b) indicate that the Hellinger distance between the output and ideal increases from 0.34\% to 7\%, and 15\% in periods 1, and 2, respectively. To examine the raw data of the obtained histograms for the non-adaptive case, refer to the crimson colored bars in Fig.~\ref{fig:output_quality}~(a).
\subsubsection{Adaptive estimation}
\begin{table}[htbp]
\caption{Estimated Pauli coefficients for peiod 1}
\begin{center}
\begin{tabular}{c c c c}
\textrm{Pauli term [$\textrm{qubit} 0 \otimes \textrm{qubit} 1$]} & \textrm{Estimated value} & \textrm{True value}&\\
\toprule
II&0.311&0.326&\\
IX&0.064&0.064&\\
IY&0.064&0.064&\\
IZ&0.097&0.078&\\
XI&0.078&0.083&\\
XX&0.016&0.016&\\
XY&0.016&0.016&\\
XZ&0.024&0.02&\\
YI&0.078&0.083&\\
YX&0.016&0.016&\\
YY&0.016&0.016&\\
YZ&0.024&0.02&\\
ZI&0.114&0.12&\\
ZX&0.023&0.024&\\
ZY&0.023&0.024&\\
ZZ&0.036&0.029&\\
\bottomrule
\end{tabular}
\end{center}
\label{tab:coefficients2}
\end{table}
\begin{table}[htbp]
\caption{Estimated coefficients for period 2}
\begin{center}
\begin{tabular}{c c c c}
\textrm{Pauli term [$\textrm{qubit} 0 \otimes \textrm{qubit} 1$]} & \textrm{Estimated value} & \textrm{True value}&\\
\toprule
II&0.243&0.26&\\
IX&0.073&0.075&\\
IY&0.073&0.075&\\
IZ&0.103&0.079&\\
XI&0.074&0.082&\\
XX&0.022&0.024&\\
XY&0.022&0.024&\\
XZ&0.031&0.025&\\
YI&0.074&0.082&\\
YX&0.022&0.024&\\
YY&0.022&0.024&\\
YZ&0.031&0.025&\\
ZI&0.103&0.108&\\
ZX&0.031&0.031&\\
ZY&0.031&0.031&\\
ZZ&0.044&0.033&\\
\bottomrule
\end{tabular}
\end{center}
\label{tab:coefficients3}
\end{table}

\begin{figure*}[!h]
\centering
\begin{tabular}{ c @{\hspace{40pt}} c }
\includegraphics[width=.84\columnwidth]{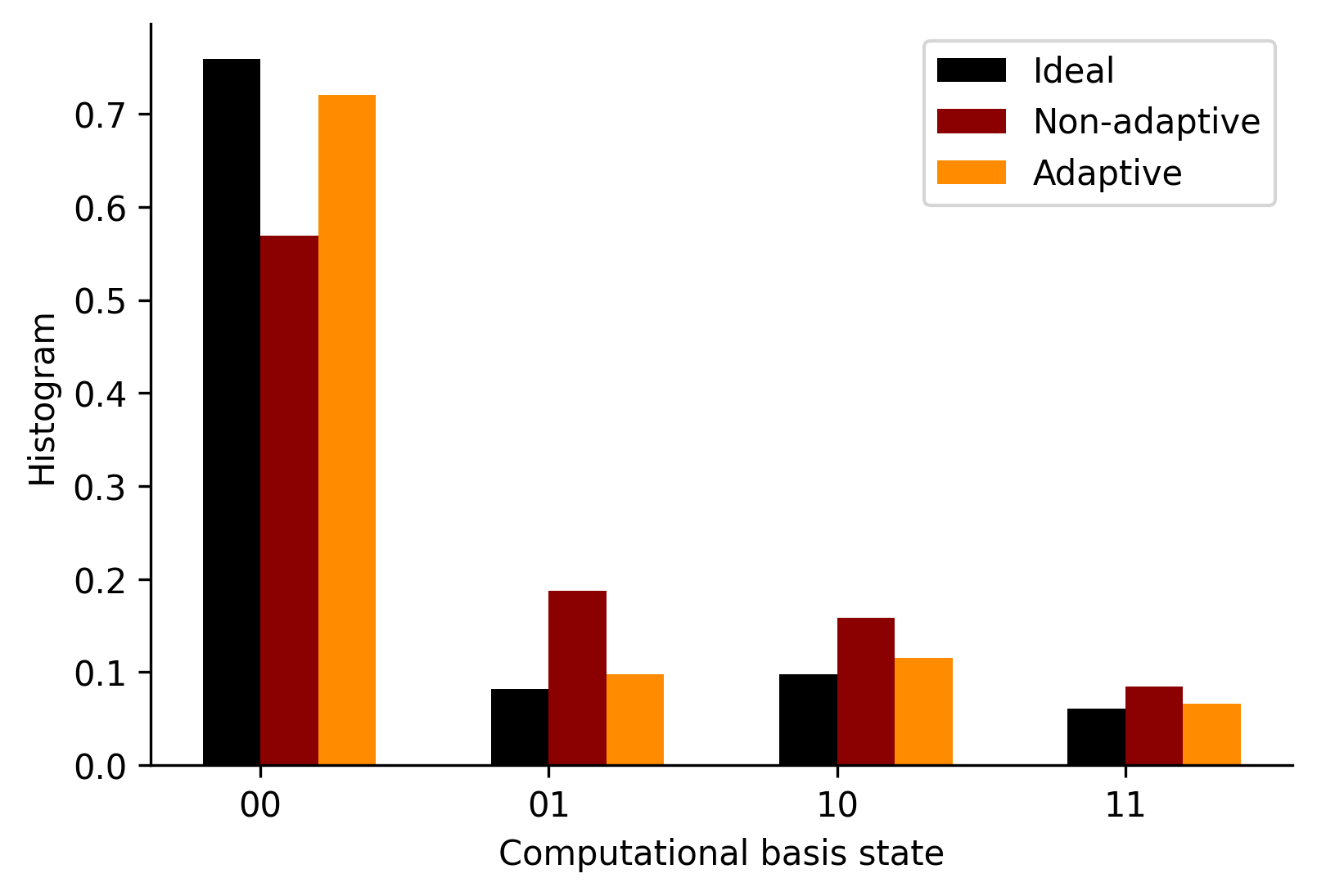}&
\includegraphics[width=.84\columnwidth]{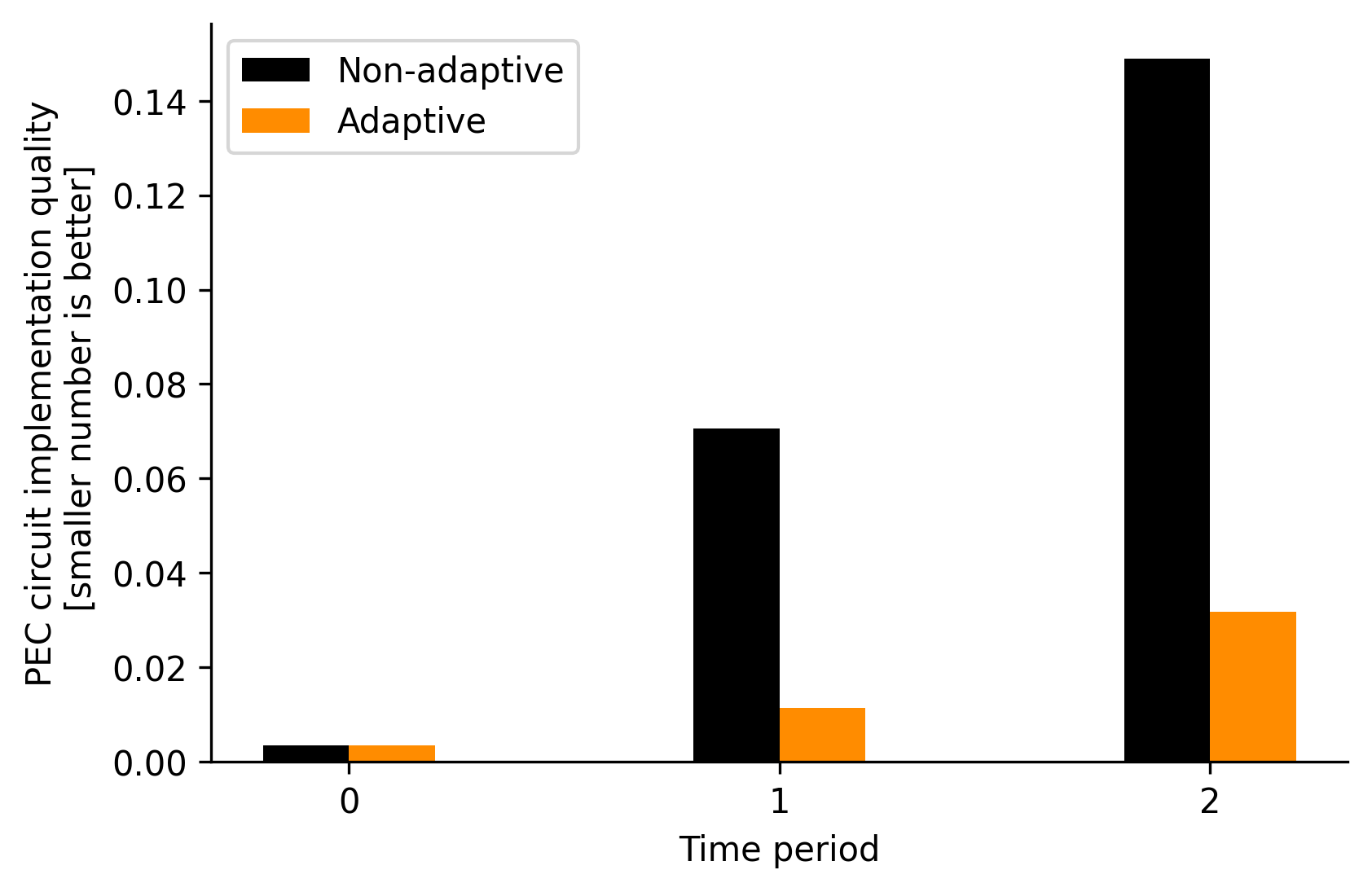}\\
\small (a) & \small (b)\\
\end{tabular}
\medskip
\caption{(a) This graph compares adaptive and non-adaptive PEC for the $H\otimes H$ gate under time-varying noise. The y-axis shows the probability of observing a basis state. The black bars represent the ideal histogram for the test input from Eq.~(\ref{eq:random_input}), while red/orange bars are non-adaptive PEC results. Adaptive PEC improves accuracy. For example, the $\ket{00}$ probability increases from 57\% to 72\%. 
(b) This graph compares adaptive and non-adaptive PEC implementations over four time-periods with time-varying noise, using Hellinger distance to measure the difference from the ideal distribution. The black and orange bars represent the observed distributions with non-adaptive and adaptive PEC, respectively. Adaptive PEC significantly outperforms non-adaptive PEC in reducing Hellinger distance to 1.1\%, and 3.2\% compared to 7\%, and 15\% for non-adaptive PEC across time-periods. Adaptive PEC uses adaptive estimation of noise super-operators to improve accuracy compared to non-adaptive PEC. Time-varying noise underscores the need for adaptive PEC.}
\label{fig:output_quality}
\end{figure*}
To improve the accuracy of the PEC-based circuit implementation in the presence of time-varying noise characteristics, we utilize Bayesian inference for dynamic channel estimation. Bayesian inference can require fewer samples than maximum likelihood to produce reliable parameter estimates, provided that the prior specification is of high quality and the model complexity is not too high.
\par 
For the Bayesian update, we use the histogram of projective measurements obtained from applying the old PEC circuit to a test density matrix, as shown in Eq.~(\ref{eq:random_input}). This test density matrix represents a known 2-qubit quantum state that produces an uneven probability distribution, which is important for accurately estimating the Pauli coefficients later. The resulting stream of 2-bit strings belongs to one of four possibilities: ${00, 01, 10, 11}$ with corresponding probabilities as 0.76, 0.08, 0.10, and 0.06, respectively. The test density matrix can be replaced with any reliably prepared quantum state that has sufficient off-diagonal components to produce a rich histogram.
\begin{equation}
\begin{aligned}
&\rho_\textrm{test} =\\
& \small{\begin{pmatrix}
0.2&0.22-0.02j&0.15-0.09j&0.16-0.1j\\
0.22+0.02j&0.24&0.16-0.08j&0.19-0.1j\\
0.15+0.09j&0.16+0.08j&0.36&0.14+0.06j\\
0.16+0.1j&0.19+0.1j&0.14-0.06j&	0.21\\
\end{pmatrix} \; .}
\end{aligned}
\label{eq:random_input}
\end{equation}
\par 
Fig.~\ref{fig:output_quality}~(b) compares the performance of adaptive and non-adaptive PEC implementations across four back-to-back time-periods in the presence of time-varying noise. The y-axis represents the Hellinger distance, which is the distance between two discrete probability distributions over the computational basis states. The black bars show the experimentally observed distribution when non-adaptive PEC is used, while the orange bars show the distribution when adaptive PEC is used. The x-axis represents the four time-periods. When using non-adaptive PEC, the Hellinger distance from the ideal distribution is 7\%, and 15\% for time-periods 1, and 2, respectively. When using adaptive PEC, the Hellinger distance significantly improves to 1.1\%, and 3.1\% for the same time-periods. This graph demonstrates that adaptive PEC improves the accuracy of the PEC circuit implementation in the presence of time-varying noise compared to non-adaptive PEC. The improvement can be attributed to the fact that the noise characteristics of the Pauli channel change over time, becoming increasingly worse. As a result, the previously implemented PEC circuits become inaccurate since the super-operators characterizing the noisy basis circuits are no longer valid. Without adaptive mitigation, two issues arise: the old linear combination coefficients for importance sampling of the Monte-Carlo circuits become invalid, and new types of noise during execution affect the circuits post-generation. This underscores the need for adaptive PEC, which uses adaptive estimation of the noise super-operators to address these issues and improve accuracy in the presence of time-varying noise.
\par 
The posterior distribution, a 16-dimensional Dirichlet distribution over the Pauli noise channel coefficients, is used to estimate the MAP (maximum-a-posteriori) of the coefficients. Table~\ref{tab:coefficients2} and Table~\ref{tab:coefficients3} demonstrate the MAP estimation performance at granularity of individual Pauli coefficients and illustrates the quality of Bayesian estimation.  With this new knowledge, we estimate the super-operators for the noisy basis circuits and the linear combination of these super-operators to design the actual circuits. The PEC is then implemented using the estimated super-operators, resulting in improved output quality, as shown in Fig.~\ref{fig:output_quality}~(a) which compares the output distribution over computational basis states for adaptive and non-adaptive PEC in the presence of time-varying noise. In this figure, the y-axis represents the probability of observing a particular computational basis state. In this two-qubit case, there are four possible states, and their probabilities  sum to 1. The black bars indicate the ideal probabilities, while the red and orange bars represent the probabilities obtained with non-adaptive and adaptive PEC, respectively, for the fourth time-period. The graph demonstrates that adaptive PEC, which used adaptive estimation of the noise super-operators, improves accuracy compared to non-adaptive PEC. Specifically, the probability of observing 00 in period 2 increases from 57\% for non-adaptive PEC to 72\% for adaptive PEC in the presence of time-varying noise. 

Note that we used a Pauli noise channel with all $4^n$ terms. However, in practical applications, it becomes necessary to reduce the number of terms. To achieve this, one can explore the use of a sparse Lindbladian noise model \cite{van2023probabilistic}, which considers noise only in nearest-neighbor connections for Pauli terms with weight greater than 1. This reduction in terms leads to a linear scaling instead of exponential with the number of qubits, making the model more computationally efficient. Also note that while it has been observed that single-qubit gate noise can be more than 10 times smaller than two-qubit gate noise \cite{van2023probabilistic}, it cannot be disregarded in Probabilistic Error Cancellation (PEC) due to error propagation effects. Moreover, in the presence of time-varying quantum noise, it becomes even more critical to account for single-qubit noise to ensure accurate error cancellation.

Note that recalibrating the noise model does not solve the challenge of dynamic estimation, as experimental evidence from various studies indicates significant fluctuations in decoherence times over time. These fluctuations, observed in studies like \cite{carroll2021dynamics, mcrae2021reproducible}, show that decoherence times can vary by approximately 50\% within an hour due to the presence of oxides on superconductors' surfaces, represented as fluctuating two-level systems (TLS) \cite{muller2015interacting, klimov2018fluctuations}. These fluctuations at the scale of minutes and hours are considered non-systematic noise, which cannot be addressed solely through re-calibrations. This emphasizes the necessity for time-varying adaptive algorithms. In contrast to methods focused on re-calibration, our approach is more versatile, applicable to situations where incomplete knowledge of the noise parameters (e.g. Pauli weights) is unavoidable, regardless of the source of imperfection or inaccuracy, be it systematic or non-systematic noise.

\section{Conclusion}\label{sec:conc}
In this study, we explore the spatio-temporal stochasticity of the Pauli noise channel, which captures the non-stationarity of the quantum noise in superconducting qubits. We present the variation of the coefficients of the noise channel for IBM Belem. We use a Dirichlet distribution to model the joint distribution of the coefficients of the Pauli noise channel. 
The Hellinger distance between two time-varying Dirichlet distributions quantifies the degree of non-stationarity, which is used to measure reliability of error channel characterization. 
We characterize the effect of the time-varying Pauli noise on the quantum information encoded in an n-qubit register and show how to obtain the coefficients of a separable 2-qubit Pauli noise channel. These coefficients are directly linked to the decoherence times of individual register elements, which exhibit stochastic fluctuations in time and across the register, leading to strong correlations among the Pauli coefficients. 
\par 
The probabilistic error cancellation (PEC) technique of quantum error mitigation provides an ideal circuit in the asymptotic Monte Carlo circuit limit if the noise is perfectly characterized. However, in the presence of non-stationary time-varying noise, an adaptive approach becomes necessary. We propose an adaptive Bayesian inference approach to improve the performance of PEC for an unreliable platform affected by time-varying noise. The time-varying Dirichlet distribution of the Pauli coefficients is dynamically estimated using a Bayesian inference-based rolling update. We present an application of PEC for executing a Hadamard operation on two-qubits in the presence of time-varying noise using adaptive estimation. The results of the simulation indicate that the previously implemented PEC approach becomes inaccurate with the change in the noise characteristics of the Pauli channel. Without adaptive mitigation, two issues arise: the old linear combination coefficients for importance sampling of the Monte-Carlo circuits are no longer valid, and the circuits, post-generation, are impacted by new types of noise during execution. The output implementation is severely affected. Overall, this study provides important insights into the behavior of non-stationary noise channels based on real-world quantum computing platforms.
\par
In this study, we considered the spatial correlations for multi-qubit noise by treating the latter as a separable collection of single-qubit channels, while retaining the spatial correlations between the different single-qubit $T_{1}$ and $T_{2}$ times. A more general approach would be to consider non-separable multi-qubit noise channels, with collective $T_{1}$ and $T_{2}$ times. This will allow us to capture non-classical correlations between qubits due to the entangling nature of multi-qubit noise and understand their effects on the performance of the PEC method. The focus of this work was on the adaptive implementation of PEC. However, the quantum estimation procedure that is implicitly involved in this discussion could be addressed in more detail. This work utilized adaptive algorithms with Bayesian inference and a Dirichlet prior to achieve probabilistic error cancellation in the presence of time-varying quantum noise. Real-world noise data from quantum devices was used for simulations. Future research may explore the implications of metrological bounds \cite{danageozian2022noisy} on multi-parameter quantum estimation using quantum Fisher information techniques for practical applications. Additionally, future investigations aim to validate and demonstrate the algorithm's efficacy with an actual quantum device.

\par The conventional PEC method faces challenges in scalability due to its preference for a complete noise characterization to achieve maximum accuracy \cite{temme2017error}. To address this, some suggestions have been made to enhance tractability by reducing the number of parameters that require estimation. One approach involves reducing overhead in quasi-probability representation construction by omitting terms dependent on the noise coefficient magnitude. This approach allows for better scalability but comes at the expense of accuracy, leaving the user to decide their tolerance for inaccuracy. Another suggested approach, as presented in \cite{van2023probabilistic}, increases reproducibility by removing the user's freedom to set the inaccuracy tolerance. It involves using sparse Lindbladian models, assuming only qubits with physical connections experience noise, leading to linear resource estimation costs instead of exponential. The adaptive approach we propose is model-agnostic and can be used with sparse Lindbladian models. The scalability, therefore, depends on the nature of the effective model deployed, with our approach being blind to these specifics.

\par Noise characterizations are assumed to be constant in current research in quantum computing when applying error mitigation. Thus, a natural question arises as to how sensitive the output is to variable noise processes. In this study, we have used an adaptive scheme using Bayesian inference to dynamically estimate the noise channels, and used that knowledge to compensate the error. This helps stabilize the circuit in terms of minimizing fluctuations with respect to the time-varying noise. The algorithm does not require a-priori knowledge of channel calibration. In fact, erroneous channel calibration knowledge is acceptable too - as the method learns dynamically from the measurements (binary strings) observed. Our work fills a gap for utilizing unstable quantum platforms with the goal of improving reproducibility in the burgeoning field of quantum information science.

%
%
\section*{ACKNOWLEDGMENTS}
\small{We thank the anonymous reviewers for their helpful feedback. This work is supported by the U.~S.~Department of Energy (DOE), Office of Science, National Quantum Information Science Research Centers, Quantum Science Center and the Advanced Scientific Computing Research, Advanced Research for Quantum Computing program, and the U.S. Army Research Office through the U.S. MURI Grant No. W911NF-18-1-0218. This research used computing resources of the Oak Ridge Leadership Computing Facility, which is a DOE Office of Science User Facility supported under Contract DE-AC05-00OR22725. The manuscript is authored by UT-Battelle, LLC under Contract No.~DE-AC05-00OR22725 with the U.S. Department of Energy. The U.S.~Government retains for itself, and others acting on its behalf, a paid-up nonexclusive, irrevocable worldwide license in said article to reproduce, prepare derivative works, distribute copies to the public, and perform publicly and display publicly, by or on behalf of the Government.  
The Department of Energy will provide public access to these results of federally sponsored research in accordance with the DOE Public Access Plan: http://energy.gov/downloads/doe-public-access-plan.}

\bibliographystyle{unsrt}
\bibliography{references.bib}
\end{document}